\documentclass[prl,10pt,twocolumn]{revtex4}
\usepackage{amsmath}
\usepackage{amssymb}
\usepackage{graphics,graphicx}

\input{tcilatex}

\begin{document}

\title{The Nature of Asymmetry in Fluid Criticality }
\author{Mikhail A. Anisimov* and Jingtao Wang}
\affiliation{Department of Chemical \& Biomolecular Engineering and Institute for
Physical Science \& Technology, University of Maryland, College Park,
Maryland 20742, USA}
\date{\today }

\begin{abstract}
By combining accurate liquid-vapor coexistence and heat-capacity data, we
have unambiguously separated two non-analytical contributions of liquid-gas
asymmetry in fluid criticality and showed the validity of "complete scaling"
[Fisher \textit{et al}., Phys. Rev. Lett. \textbf{85}, 696 (2000); Phys. \
Rev. E,\textit{\ }\textbf{67}, 061506, (2003)]. We have also developed a
method to obtain two scaling-field coefficients, responsible for the two
sources of the asymmetry, from mean-field equations of state. Since the
asymmetry effects are completely determined by Ising critical exponents,
there is no practical need for\ a special renormalization-group theoretical
treatment of asymmetric fluid criticality.

*anisimov@umd.edu
\end{abstract}

\maketitle

\address{Department of Chemical and Biomolecular Engineering and Institute\\
for Physical Science and Technology, University of Maryland, College Park,\\
MD 20742}

A controversial issue of liquid-gas asymmetry in fluids has been a subject
of prolonged discussions for more than a century since Cailletet and
Mathias' discovery of the empirical "law"\ of rectilinear diameter \cite%
{Cailletet:ASCRH1886}. According to this "law", the mean of the densities of
liquid $\rho ^{\prime }$ and saturated vapor $\rho ^{\prime \prime }$ is a
linear function of the temperature $T$:

\begin{equation}
\Delta \hat{\rho}_{\text{d}}\equiv \frac{\rho ^{\prime }+\rho ^{\prime
\prime }}{2\rho _{\text{c}}}=1+D\left\vert \Delta \hat{T}\right\vert ,
\end{equation}%
where $\rho _{\text{c}}$ is the critical density, $\Delta \hat{T}=(T-T_{%
\text{c}})/T_{\text{c}}$ the reduced distance to the critical temperature $%
T_{\text{c}}.$ The system-dependent coefficient $D$ generally increases from
0.02 for $^{3}$He \cite{Hahn:JLTP2004} to values larger than unity with
increase of $T_{\text{c}}$ \cite{Pestak:PRB1987,Singh:JCP1990}. In this
Letter we demonstrate how the meanfield rectilinear diameter splits up in
the critical region into two "singular diameters" associated with two
different sources of asymmetry. It is commonly accepted that fluids
asymptotically $(\Delta \hat{T}\rightarrow 0)$ belong to the critical-point
universality class of the Ising model \cite{F:82}. We argue that the
asymmetry effects in near-critical fluids, at least in lower approximation,
are determined by Ising critical exponents, hence, in contrast to a commonly
used approach, there is no need for\ a special renormalization-group
theoretical treatment of asymmetric fluid criticality.

Thermodynamics near a critical point is controlled by two scaling fields,
\textquotedblleft ordering\textquotedblright\ $h_{1}$ and \textquotedblleft
thermal\textquotedblright\ $h_{2},$ while an appropriate field-dependent
potential $h_{3}$ is a universal function of $h_{1}$ and $h_{2}$ \cite{F:82}:

\begin{equation}
h_{3}=h_{2}^{2-\alpha }f^{\pm }\left( \frac{h_{1}}{h_{2}^{\beta +\gamma }}%
\right) ,
\end{equation}%
where \ $\alpha =0.109$, $\beta =0.326,$ and $\gamma =1.239,$ interrelated
as $\alpha +2\beta +\gamma =2$, are universal Ising critical exponents in
the scaling power laws (as a function of $h_{2}$ at $h_{1}=0$) for the
weakly-divergent susceptibility, order parameter, and strongly-divergent
susceptibility, respectively \cite{F:82}. The scaling "densities", a
strongly fluctuating order parameter $\phi _{1}$ and a weakly fluctuating $%
\phi _{2},$ conjugate to $h_{1}$ and $h_{2}$, such that $dh_{3}=\phi
_{1}dh_{1}+\phi _{2}dh_{2}.$ The universal function $f^{\pm }$ contains two
system-dependent amplitudes, and the superscript $\pm $ refers to $%
h_{2}\gtrless 0.$ The Ising model formulated for fluids is known as
"lattice-gas" \cite{Yang:PRL1964}. In the lattice-gas $h_{3}$ is the
"critical part" of the grand thermodynamic potential, $\Omega =-PV,$ taken
per unit volume, thus $h_{1}$ is the dimensionless chemical-potential
difference $\Delta \hat{\mu}=(\mu -\mu _{\text{c}})/k_{\text{B}}T_{\text{c}}$%
, where $\mu _{\text{c}}$ is the value of the chemical potential at the
critical point and $k_{\text{B}}$ is Boltzmann's constant. By definition, $%
h_{1}=0$ along the critical isochore above $T_{\text{c}}$ and along the
liquid-vapor coexistence curve below $T_{\text{c}}$. Hence, in the lattice
gas the order parameter $\phi _{1}=\Delta \hat{\rho}=(\rho -\rho _{\text{c}%
})/\rho _{\text{c}}$, the thermal scaling field $h_{2}=\Delta \hat{T}$, and $%
\phi _{2}=\Delta (\hat{\rho}\hat{S})=(\hat{\rho}\hat{S}-\hat{\rho}_{c}\hat{S}%
_{c})/k_{\text{B}},$ where $\hat{\rho}\hat{S}$ is the density of entropy.

The lattice gas has a special symmetry: the order parameter is symmetric
with respect to the critical isochore ($D=0$). Since the early 1970's the
liquid-gas asymmetry has been commonly incorporated into the lattice-gas
analogy by linear mixing of two independent physical fields $\Delta \hat{\mu}
$ and $\Delta \hat{T}$ into the both theoretical scaling fields $h_{1}$ and $%
h_{2}$ \cite{Mermin}. Since the absolute value of entropy is arbitrary,
mixing of $\Delta \hat{T}$ into $h_{1}$ plays no role. Contrarily, mixing of 
$\Delta \hat{\mu}$ (with $b_{2}$ as a mixing coefficient) into $h_{2}$ has
an important consequence, known as the "singular diameter": the mean of the
densities must contain a non-analytic contribution $\varpropto
b_{2}\left\vert \Delta \hat{T}\right\vert ^{1-\alpha }$, so that $d(\Delta 
\hat{\rho}_{\text{d}})/d\hat{T}\varpropto b_{2}\left\vert \Delta \hat{T}%
\right\vert ^{-\alpha }$, diverging weakly. However, the chemical potential
would remain an analytical function of temperature along the liquid-vapor
coexistence and the symmetry would be restored by a redefinition of the
order parameter as $\phi _{1}=\Delta \hat{\rho}+b_{2}\Delta (\hat{\rho}\hat{S%
}).$

At this point we encounter a\ major problem. \ First of all, the existence
of the $\left\vert \Delta \hat{T}\right\vert ^{1-\alpha }$ term in the
"diameter" of real fluids has never been detected unambiguously . While some
fluids show strong deviations from rectilinear diameter, apparently even
stronger than $\left\vert \Delta \hat{T}\right\vert ^{1-\alpha }$\cite%
{Pestak:PRB1987}, many fluids show very little or no deviations at all \cite%
{Widom:book1972,Narger:PRB1990}. Moreover, there is a conceptual problem
with mapping real fluids into the lattice-gas even at the mean-field level.
In the mean-field approximation the critical part of the thermodynamic
potential is represented by a Landau expansion:

\begin{equation}
h_{3}=\frac{1}{2}a_{0}h_{2}\phi _{1}^{2}+\frac{1}{4}u_{0}\phi
_{1}^{4}-h_{1}\phi _{1}.
\end{equation}%
When $h_{1}=\Delta \hat{\mu},$ $h_{2}=\Delta \hat{T}$ $+b_{2}\Delta \hat{\mu}%
,$ and $\phi _{1}=\Delta \hat{\rho}+b_{2}\Delta (\hat{\rho}\hat{S}),$ this
expansion generates asymmetric terms $\varpropto b_{2}\Delta \hat{T}(\Delta 
\hat{\rho})^{3}$ and $\varpropto b_{2}(\Delta \hat{\rho})^{5}.$ However, in
the simplest equation of state that describes real-fluid behavior, the van
der Waals equation, the term $\varpropto \Delta \hat{T}(\Delta \hat{\rho}%
)^{3}$ is absent, while the term $\varpropto (\Delta \hat{\rho})^{5}$
exists. Furthermore, in most classical equations of state $d\hat{\mu}^{2}/d%
\hat{T}^{2}$ along the liquid-vapor coexistence exhibits a discontinuity
directly related to the existence of the independent 5th-order term \cite%
{N:81} in Landau expansion. The existence of the independent 5th-order term
makes exact mapping of fluids into the lattice-gas model by the conventional
mixing of physical fields impossible. On the other hand, a
renormalization-group treatment of the 5th-order term resulted in an
independent critical exponent $\theta _{5}\simeq 1.3$\cite{Zhang:book1983}
(not existing in the Ising model!).\ We show, however, that asymmetric
fluids can be consistently mapped into Ising criticality by applying
so-called "complete scaling" originally proposed by Fisher \textit{et al.} 
\cite{Fisher:PRL2000}. A redefinition of the order parameter, suggested by
complete scaling, results in elimination of the 5th-order term in Landau
expansion, thus making the renormalization-group treatment of the 5th-order
term irrelevant for fluids.

Complete scaling suggests that all three physical fields $\Delta \hat{\mu},$ 
$\Delta \hat{T},$ and $\Delta \hat{P}=(P-P_{\text{c}})/\rho _{\text{c}}k_{%
\text{B}}T_{\text{c}}$ are equally mixed into three scaling fields $h_{1},$ $%
h_{2},$ and $h_{3}$. In linear approximation

\begin{eqnarray}
h_{1} &=&a_{1}\Delta \hat{\mu}+a_{2}\Delta \hat{T}+a_{3}\Delta \hat{P},\text{
} \\
h_{2} &=&b_{1}\Delta \hat{T}+b_{2}\Delta \hat{\mu}+b_{3}\Delta \hat{P}, \\
h_{3} &=&c_{1}\Delta \hat{P}+c_{2}\Delta \hat{\mu}+c_{3}\Delta \hat{T}.
\end{eqnarray}%
Before we apply complete scaling to describe asymmetry in fluids, we note
that the relations between scaling and physical fields can be simplified.
The coefficients $a_{1}$ and $b_{1}$ can be absorbed in two amplitudes in
the scaling function $f^{\pm }.$ The coefficients $c_{1}$ and $c_{2}$ are
absorbed by making the thermodynamic potential $h_{3}$ dimensional. The
coefficient $c_{3}=\hat{S}_{\text{c}}$ is determined by the choice of \ the
critical value of entropy. By adopting $\hat{S}_{\text{c}}=\left( k_{\text{B}%
}\rho _{\text{c}}\right) ^{-1}\left( \partial P/\partial T\right) _{h_{1}=0,%
\text{c}}=\left( d\hat{P}/d\hat{T}\right) _{\text{cxc},\text{c}},$ the slope
of the saturation-pressure curve at the critical point, one obtains $%
a_{2}=-a_{3}\left( d\hat{P}/d\hat{T}\right) _{\text{cxc},\text{c}}$ and $%
b_{3}=0$. Finally, the scaling fields contain only two amplitudes
responsible for asymmetry in fluid criticality:

\begin{eqnarray}
h_{1} &=&\Delta \hat{\mu}+a_{3}[\Delta \hat{P}-\left( d\hat{P}/d\hat{T}%
\right) _{\text{cxc},\text{c}}\Delta \hat{T}],\text{ } \\
h_{2} &=&\Delta \hat{T}+b_{2}\Delta \hat{\mu},\text{ } \\
h_{3} &=&-\Delta \hat{P}+\Delta \hat{\mu}+\left( d\hat{P}/d\hat{T}\right) _{%
\text{cxc},\text{c}}\Delta \hat{T}.
\end{eqnarray}%
As a result, while the order parameter in fluids is, in general, a nonlinear
combination of density and entropy $\phi _{1}=[\Delta \hat{\rho}+b_{2}\Delta
(\hat{\rho}\hat{S})]/(1+a_{3}\Delta \hat{\rho}),$ the weakly fluctuating
scaling density $\phi _{2}$ in first approximation is associated with the
density of entropy only, $\phi _{2}=\Delta (\hat{\rho}\hat{S})$. There are
two important thermodynamic consequences of complete scaling\ that can be
checked experimentally. Firstly, the \textquotedblleft
diameter\textquotedblright\ $\rho _{\text{d}}$ should contain two
non-analytical contributions, associated with the terms $a_{3}\Delta \hat{P}$
and $b_{2}\Delta \hat{\mu}$ in the scaling fields:

\begin{align}
\hat{\rho}_{\text{d}}-1& =a_{3}\left( 1+a_{3}\right) \phi _{1}^{2}+b_{2}\phi
_{2}+\ldots  \notag \\
& =D_{1}\left\vert \Delta \hat{T}\right\vert ^{2\beta }+D_{2}\left\vert
\Delta \hat{T}\right\vert ^{1-\alpha }+D_{3}\left\vert \Delta \hat{T}%
\right\vert +\ldots
\end{align}%
where $D_{1}=a_{3}B_{0}^{2}/(1+a_{3})$ and $D_{2}=b_{2}A_{0}^{-}/(1-\alpha )$
with $B_{0}$ and $A_{0}^{-}$ being the amplitudes in the asymptotic scaling
power laws for the liquid/vapor densities, $\Delta \hat{\rho}=\pm
B_{0}\left\vert \Delta \hat{T}\right\vert ^{\beta }+...,$ and isochoric heat
capacity in the two-phase region, $C_{V}/k_{\text{B}}=A_{0}^{-}\left\vert
\Delta \hat{T}\right\vert ^{-\alpha }+...$ $(A_{0}^{-}=A_{0}^{+}/0.523)$
[5c]. Note, since $2\beta <1-\alpha ,$ the term $D_{1}\left\vert \Delta \hat{%
T}\right\vert ^{2\beta }$ should dominate near the critical point .
Secondly, the presence of this term implies a so-called Yang-Yang anomaly:
the divergence of the heat capacity in the two-phase region is shared among
the second derivatives of pressure and chemical potential \cite%
{Fisher:PRL2000}. Experimental verification of complete scaling\ is a very
challenging task. The nonanalytical contributions in the \textquotedblleft
diameter\textquotedblright\ are usually not large enough to be separated
unambiguously. Attempts to fit some experimental and simulation data to Eq.
(10) showed very poor conversions \cite{Kim:CPL2005}, mainly because of a
strong correlation between the linear and $D_{2}\left\vert \Delta \hat{T}%
\right\vert ^{1-\alpha }$ terms. Experimental tests of the Yang-Yang anomaly
are even more controversial since traces of impurities can easily mimic such
an anomaly, thus making any conclusions unreliable \cite{W2002}.

We have been able to reliably determine the two asymmetry coefficients, $%
a_{3}$ and $b_{2},$ and to conclusively prove the validity of complete
scaling by combining accurate experimental and simulation liquid-vapor
coexistence and heat-capacity data. We have exploited the fact that the
coefficients $D_{2}$ and $D_{3}$ in Eq. (10) are actually coupled. As the
weakly fluctuation scaling density $\phi _{2}$ is the critical part of the
entropy density, in the two-phase region at average density $\rho =\rho _{%
\text{c}}$

\begin{equation}
\phi _{2}=\dint \frac{C_{V}^{\text{cr}}}{k_{\text{B}}T}dT=-\frac{A_{0}^{-}}{%
1-\alpha }\left\vert \Delta \hat{T}\right\vert ^{1-\alpha }+B_{\text{cr}%
}\left\vert \Delta \hat{T}\right\vert ,  \label{heatcapacity}
\end{equation}%
where $C_{V}^{\text{cr}}$ is the critical part of the isochoric heat
capacity and $B_{\text{cr}}$ the so-called "critical background",
fluctuation-induced analytical part of the heat capacity. The critical
background can be obtained from a ratio between $B_{\text{cr}}$, $A_{0}^{+}$%
, and the non-asymptotic heat capacity amplitude $A_{1}^{+}$ \cite%
{Balgnus:PRB1985} and by subtracting the "ideal gas" from the total
heat-capacity background. Both procedures yield very similar values of $B_{%
\text{cr}}.$ Since $D_{3}=-b_{2}B_{\text{cr}}$, Eq. (10) contains only two
adjustable coefficients, $a_{3}$ and $b_{2}.$ We have examined a number of
systems, real fluids and simulated models \cite%
{Pestak:PRB1987,Artyukhovskaya}, for which we could find both heat-capacity
and coexistence data in the range $\left\vert \Delta \hat{T}\right\vert
<0.01 $. In this range the terms of higher-order than linear in Eq. (10) are
within experimental errors. Experimental data closer than $\left\vert \Delta 
\hat{T}\right\vert <10^{-4}$ were avoided as they might be affected by
errors\ in $\rho _{\text{c}}$ and $T_{\text{c}}$ and by other factors, such
as gravity, impurities, \textit{etc}.. For all systems studied we have been
able to obtain reliable values of $B_{\text{cr}}$ and conclusively separate
two singular contributions to the diameter. Two typical examples that
represent two different kinds of asymmetry in fluid criticality are shown in
Fig. 1.

\begin{figure}[tp]
\begin{center}
\includegraphics*[width=0.35\textwidth,angle=-90]{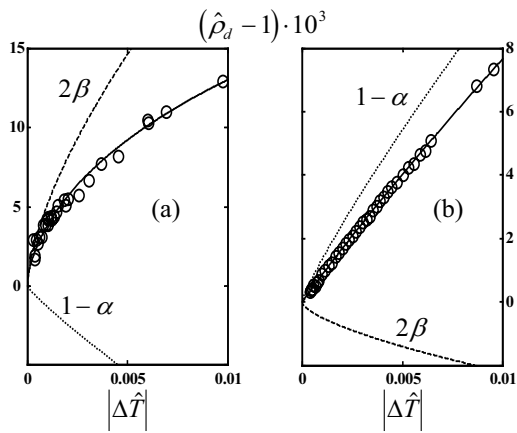}
\end{center}
\caption{Singular diameters in SF$_{6}$ (a) and N$_{2}$ (b). Experimental
data: SF$_{6}$ [16c] and N$_{2}$ \protect\cite{Pestak:PRB1987}$.$ Curves:
solid - fit to Eq. (10), dashed - $2\protect\beta $ term, dotted - $1-%
\protect\alpha $ and linear terms. Heat-capacity source [16d] (SF$_{6})$ and
[16i] (N$_{2}).$}
\label{fig1}
\end{figure}

In diameters of some fluids, such as SF$_{6},$ C$_{2}$F$_{3}$Cl$_{3},$ and 
\textit{n}-C$_{7}$H$_{16},$ the $\left\vert \Delta \hat{T}\right\vert
^{2\beta }$ term dominates ($a_{3}$ is relatively large and positive ) while
in many other fluids, such as HD, Ne, N$_{2}$, and CH$_{4},$ the two
singular contributions in diameter partially compensate each other ($a_{3}$
is small and negative), creating an illusion of rectilinear diameter even
close to the critical point. In Fig. 2 the two asymmetry coefficients are
plotted against the dimensional density $\rho ^{\ast }$ defined as $\rho
^{\ast }=\rho _{\text{c}}(8\xi _{0}^{3}),$ where $\xi _{0}$ is the amplitude
of the correlation length (representing the range of interactions) obtained
from the heat-capacity amplitude $A_{0}^{+}$ through the two-scale factor of
universality, $A_{0}^{+}\rho _{\text{c}}\xi _{0}^{3}=0.171$ [5c]. A general
trend in the two sources of asymmetry is clear: the $\left\vert \Delta \hat{T%
}\right\vert ^{2\beta }$ singularity is a dominant contribution into the
singular diameter if the ratio of critical volume per molecule $\left( \rho
_{\text{c}}^{-1}\right) $ to "interaction volume $\left( 8\xi
_{0}^{3}\right) $" is large \cite{Vink2006}.

\begin{figure}[tp]
\begin{center}
\includegraphics*[width=0.40\textwidth,angle=-90]{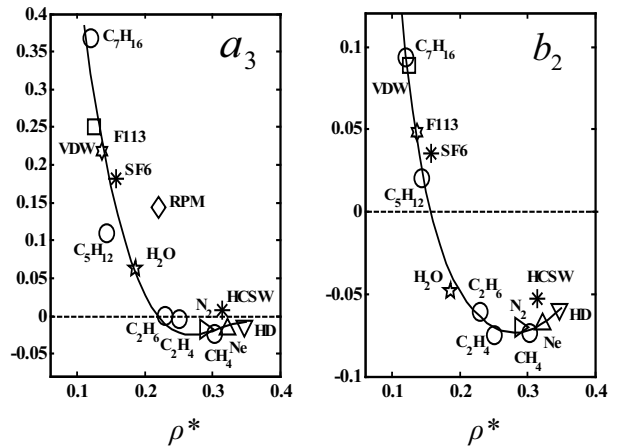}
\end{center}
\caption{Complete-scaling asymmetry coefficients $a_{3}$ and $b_{2}$ \textit{%
versus} reduced critical density $\protect\rho ^{\ast }=\protect\rho _{c}(8%
\protect\xi _{0}^{3}).$ VDW is a modified-by-fluctuations van der Waals
fluid \protect\cite{Anya2004} with a short interaction range ($R=(\protect%
\rho ^{\ast })^{1/3}=0.5$). HCSW is a simulated hard core square well model 
\protect\cite{Kim:CPL2005}. For a similated restrictive primitive model
(RPM) \protect\cite{Kim:CPL2005} $a_{3}=0.14$ and $b_{2}=-0.48$ (off scale)
with $\protect\rho ^{\ast }\simeq 0.22.$ The solid curves are given as a
guidance.}
\label{fig2}
\end{figure}

\begin{figure}[tp]
\begin{center}
\includegraphics*[width=0.35\textwidth,angle=-90]{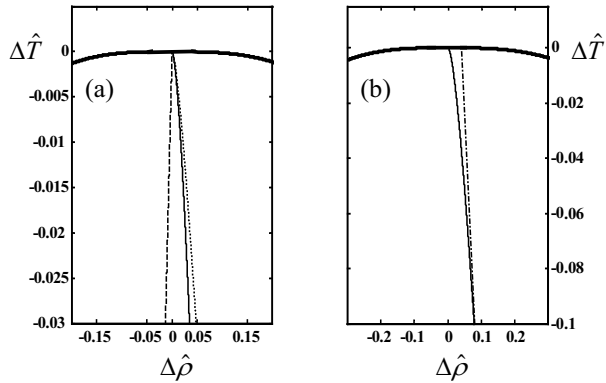}
\end{center}
\caption{Crossover diameter in a van der Waals equation of state modified by
fluctuations with a short interaction range $R=(\protect\rho ^{\ast
})^{1/3}=0.5$. Thick solid curves are the phase boundary. (a) two
contributions in the singular diameter (solid line): $1-\protect\alpha $ and
linear term (dashed line) and $2\protect\beta $ (dotted line); (b) Crossover
between rectilinear diameter (dashed-dotted line) and singular diameter
(solid line) in a broder critical region.}
\label{fig3}
\end{figure}

Assuming that the relations between physical fields and scaling fields are
not affected by fluctuations, we have developed a method to obtain the
asymmetry coefficients $a_{3}$ and $b_{2}$ from mean-field ("classical")
equations of state. By combining Eqs. (3) and (7-9), we obtain

\begin{equation}
\frac{a_{3}}{1+a_{3}}=\frac{2\mu _{21}}{3\mu _{11}}-\frac{\mu _{40}}{5\mu
_{30}},\text{ }b_{2}=\frac{\mu _{21}}{\mu _{11}^{2}}-\frac{\mu _{40}}{5\mu
_{30\mu 11}}
\end{equation}%
with $\mu _{ij}=\partial ^{i+j}\mu /\partial \hat{\rho}^{i}\partial T^{j}.$
We have obtained $a_{3}$ and $b_{2}$ for a few classical equations of state,
the fine-lattice discretization model (crossover between the van der Waals
fluid and lattice gas) \cite{Sarvin:JPC2005}, the Debye-H\"{u}ckel and
Flory-Huggins models. The coefficient of rectilinear diameter

\begin{equation}
D=\frac{a_{3}}{1+a_{3}}\frac{6\mu _{11}}{\mu _{30}}-b_{2}\frac{3\mu _{11}^{2}%
}{\mu _{30}}=\frac{a_{3}}{1+a_{3}}\bar{B}_{0}^{2}-b_{2}\frac{\Delta \bar{C}%
_{V}}{k_{\text{B}}},
\end{equation}%
where $\bar{B}_{0}$ and $\Delta \bar{C}_{V}$ are mean-field amplitudes of
coexistence densities and heat-capacity discontinuity. Close to the critical
point the rectilinear diameter, affected by fluctuations, splits into two
nonanalytical terms, shifting the critical density \ In Fig. 3 crossover
between rectilinear diameter and complete-scaling singular diameter is shown
for the van der Waals equation of state renormalized by fluctuations \cite%
{Anya2004}. A fluctuation shift in the van der Waals critical density is
mainly controlled by the $\left\vert \Delta \hat{T}\right\vert ^{2\beta }$
singularity since the van der Waals value of $a_{3}$ is relatively large ($%
a_{3}/(1+a_{3})=b_{2}/\hat{\mu}_{11}=0.2$). In the same fashion we have
calculated crossover between the discontinuity $\Delta \left( d^{2}\hat{\mu}%
/d\hat{T}^{2}\right) _{\text{cxc}}=(\mu _{11}^{2}/\mu _{30})[(-\mu _{21}/\mu
_{11})+(3\mu _{40}/5\mu _{30})]=-a_{3}/(1+a_{3})(\Delta \bar{C}_{V}/k_{\text{%
B}}),$ and a divergence known as the Yang-Yang anomaly, $\left( d^{2}\hat{\mu%
}/d\hat{T}^{2}\right) _{\text{cxc}}=a_{3}\left( d^{2}\hat{P}/d\hat{T}%
^{2}\right) _{\text{cxc}}=-a_{3}/(1+a_{3})A_{0}^{-}\left\vert \Delta \hat{T}%
\right\vert ^{-\alpha }.$ A renormalization-group treatment of the $(\Delta 
\hat{\rho})^{5}$ term instead predicts a cusp, containing a term $\varpropto
\left\vert \Delta \hat{T}\right\vert ^{\theta _{5}-\alpha -\beta }\sim
\left\vert \Delta \hat{T}\right\vert ^{0.865}$ \cite{N:81} . Similarly,
there should be a term $\propto a_{3}/(1+a_{3})\left\vert \Delta \hat{T}%
\right\vert ^{\beta -\nu }$ instead of $\left\vert \Delta \hat{T}\right\vert
^{\theta _{5}-\nu }$ \cite{Matthew Fisher} in the so-called Tolman's length,
a curvature correction to the surface tension of a liquid droplet.

We conclude that the asymmetry in near-critical fluids originates from two
sources: one is a coupling between the density and density of entropy,
another one is a non-linear coupling between the density and molecular
volume. Both sources can be incorporated into symmetric Ising criticality by
a proper mixing of physical fields into scaling fields.

We acknowledge valuable discussions with M. E. Fisher, Y. C. Kim, C. A.
Cerdeiri\~{n}a, J. V. Sengers, and B. Widom. We also thank I. M.
Abdulagatov, E. E. Gorodetski\u{\i}, and V. P. Voronov for providing us with
unpublished heat-capacity data for hydrocarbons. The research was supported
by NASA, Grant NAG32901.


\begin{thebibliography}{99}
\bibitem{Cailletet:ASCRH1886} L. \ Cailletet, and E. \ C. \ Mathias, Seanc.
Acad. Sci. Comp. Rend. Hebd., Paris \textbf{102}, 1202, (1886).

\bibitem{Hahn:JLTP2004} I. Hahn, M. Weilert, F. Zhong, and M. Barmatz, J.
Low Temp. Phys. \textbf{137}, 579, (2004).

\bibitem{Pestak:PRB1987} M. W. Pestak, M. H. W. Chan, J. R. deBruyn, D. A.
Balzarini, and N. W. Ashcroft, Phys. Rev. B \textbf{36}, 599, (1987).

\bibitem{Singh:JCP1990} R. R. Singh, and K. S. Pitzer, J. Chem. Phys. 
\textbf{92}, 3096, (1990).

\bibitem{F:82} a. M. E. Fisher, in \textit{Critical Phenomena}, F. J. W.
Hahne, ed. Lecture notes in Physics Vol. 186 (Springer, Berlin, 1982), p. 1.
b. R. Guida and J. Zinn-Justin, J. Phys. A \textbf{31}, 8103, (1998). c. M.
E. Fisher and S.-Y. Zinn, J. Phys. A \textbf{31}, L629 (1998).

\bibitem{Yang:PRL1964} C. N. Yang, and C. P. Yang, Phys. \ Rev. \ Lett. 
\textbf{13}, 303, (1964).

\bibitem{Mermin} N. D. Mermin and J. J. Rehr , Phys. \ Rev. Lett. \textbf{26}%
, 1155 (1971); V. L. Pokrovski\u{\i}, JETP Letters 17, 156 (1973); N. B.
Wilding and A. D. Bruce, J. Phys.: Condens. Matter \textbf{4}, 3087 (1992).

\bibitem{Widom:book1972} B. \ Widom, \textit{The R. A. Welch Foundation
Conferences on Chemical Research XVI. }Houston, (1972).

\bibitem{Narger:PRB1990} U. N\"{a}rger, and D. A. Balzarini, Phys. Rev. B 
\textbf{42}, 6651, (1990).

\bibitem{N:81} M. Ley-Koo and M. S. Green, Phys. Rev. A 23, 2650 (1981); J.
F. Nicoll and R. K. P. Zia, Phys. Rev. B \textbf{23}, 6157 (1981); F. Zhang
and R. K. P. Zia, J. Phys. A \textbf{15}, 3303 (1982); K. E. Newman and E.
K. Riedel, Phys. Rev. B 30, 6615 (1984).

\bibitem{Zhang:book1983} F. C. Zhang, Ph. D. Thesis, Virginia Polytechnic
Institute and State University, 1983.

\bibitem{Fisher:PRL2000} M. E. Fisher, and G. Orkoulas, Phys. Rev. Lett.%
\textit{\ }\textbf{85}, 696, (2000); Y. C. Kim, M. E. \ Fisher, and G.
Orkoulas, Phys. \ Rev. E,\textit{\ }\textbf{67}, 061506, (2003).

\bibitem{Kim:CPL2005} Y. C. Kim, M. E. Fisher, and E. Luijten, Phys. Rev.
Lett. \textbf{91}, 065701 (2003); Y. C. Kim and M. E. Fisher, Chem. Phys.
Lett. \textbf{414}, 185 (2005).

\bibitem{Balgnus:PRB1985} C. Bagnuls and C. Bervillier, Phys. Rev. B \textbf{%
32}, 7209 (1985).

\bibitem{W2002} A. K. Wyczalkowska, Y. C. Kim, M. A. Anisimov, and J. V.
Sengers , J. Chem. Phys. \textbf{116}, 4202 (2002).

\bibitem{Artyukhovskaya} a. (C$_{5}$H$_{12}$, cxc) L. M. Artyukhovskaya, E.
T. Shimanskaya, and Yu. I. Shimanski\u{\i}, Sov. Phys. JETP \textbf{32}, 375
(1971). b. (C$_{2}$F$_{3}$Cl$_{3}$, cxc) E. T. \ Shimanskaya, I. V.
Bezruchko, V. I. Basok, and Yu. I. \ Shimanski\u{\i}, Sov. Phys. JETP 
\textbf{53}, 139, (1981). c. (SF$_{6}$, cxc) J. Weiner, K. H. Langley, and
N. C. Ford, Jr, Phys. Rev. Lett.\textit{\ }\textbf{32}, 879, (1974). d. (SF$%
_{6}$, $C_{V}$) A. Haupt, and J. Straub, Phys. Rev. E \textbf{59}, 1795
(1999). e. (CH$_{4}$, cxc) R. Kleinrahm, and W. Wagner,\textit{\ }J. Chem.
Thermodynamics, \textbf{18}, 739 (1986). f. (H$_{2}$0, cxc and $C_{V}$) W.
Wagner, and A. Pru\ss ,\textit{\ }J. Phys. Chem. Refer. Data \textbf{31},
387 (2002). g. (C$_{7}$H$_{16}$, cxc) L. M. Artyukhovskaya, E. T.
Shimanskaya, and Yu. I. Shimanski\u{\i}, Sov. Phys. JETP \textbf{36}, 1140
(1973). h. (RPM, $C_{V}$) G. Orkoulas, M. E. Fisher, and A. Z.
Panagiotopoulos, Phys. \ Rev. \ E\textit{\ }\textbf{63}, 051507 (2001). i. (N%
$_{2}$ and Ar, $C_{V}$) A. V. Voronel V. G. Gorbunova, V. A. Smirnov, N. G.
Shmakov, and V. V. Shekochikhina, Sov. Phys. JETP \textbf{63}, 965 (1972).
j. (RPM, cxc; HCSW, cxc and $C_{V})$ Y. C. Kim, Phys. Rev. \ E 71, 051501
(2005).

\bibitem{Vink2006} This is supported by a recent simulation, R. L. C. Vink,
J. Chem. Phys. \textbf{124}, 094502 (2006), of the Widom-Rowlinson model [9]
for which $\rho _{\text{c}}\sigma ^{3}\simeq 0.75$ ($\sigma $ is the
particle diameter) and no $2\beta $ tem is detected.

\bibitem{Sarvin:JPC2005} S. Moghaddam, Y. C. Kim, and M. E. Fisher, J. Chem.
Phys. B\textit{, }\textbf{109}, 6824 (2005)

\bibitem{Anya2004} A. K. Wyczalkowska, J. V. Sengers, and M. A. Anisimov,
Physica A \textbf{334}, 482 (2004).

\bibitem{Matthew Fisher} M. P. A. Fisher and M. Wortis, Phys. Rev. B. 
\textbf{29}, 6252 (1984).
\end{thebibliography}
\end{document}